\documentclass[fleqn,usenatbib,useAMS]{mnras}
\usepackage{graphicx}	% Including figure files
\usepackage{amsmath}	% Advanced maths commands
\usepackage{amssymb}	% Extra maths symbols
\usepackage{multicol}        % Multi-column entries in tables
\usepackage{bm}		% Bold maths symbols, including upright Greek
\usepackage{pdflscape}	% Landscape pages

\usepackage{mathrsfs}

\usepackage{upgreek}
\title[perturbation study of magnetic degenerate stars]{A perturbation study of axisymmetric strongly magnetic degenerate stars : the case of super-Chandrasekhar white dwarfs}

\author[P. Bera and D. Bhattacharya]{Prasanta Bera\thanks{E-mail:pbera@iucaa.in}, Dipankar Bhattacharya\thanks{E-mail:dipankar@iucaa.in}\\
Inter University Centre for Astronomy and Astrophysics, Post Bag 4, Pune 411007, India.}

\begin{document}
\pagerange{\pageref{firstpage}--\pageref{lastpage}} \pubyear{0000}
\maketitle
\label{firstpage}
\begin{abstract}
In the presence of a strong magnetic field a stellar equilibrium configuration, aided by the Lorentz force, can support a larger mass than a non-magnetic one. This has been considered a possible explanation of the super-Chandrasekhar mass white dwarfs giving rise to over-luminous Type-Ia supernovae. We present here linear and non-linear perturbation studies of such strongly magetised configurations and show that axisymmetric configurations with poloidal or toroidal fields are unstable. The numerical evolution of the perturbations shows instability after about an Alfv\'en crossing time. This time scale is very short for the magnetically supported super-Chandrasekhar mass white dwarfs. Uniform rotation about the symmetry axis can reduce the growth rate but can not stabilize the super-massive configurations. It is concluded that long-lived super-Chandrasekhar mass white dwarfs supported by magnetic field are unlikely to occur in Nature.
\end{abstract}


\section{Introduction }

White dwarfs,  compact object supported by electron degeneracy pressure, have the well known maximum mass limit of 1.4 M$_\odot$ \citep{Chandrasekhar1931}. If accretion raises the mass above this Chandrasekhar limit, the white dwarf is no longer able to support itself against gravitational collapse, and the resulting rapid contraction leads to a  Type-Ia supernova. The characteristic mass limit sets the standard properties of the Type-Ia supernova. Recently however a few cases have been observed of over-luminous Type Ia supernovae that require white dwarfs well above this mass limit ($\mbox{$\stackrel{>}{_{\sim}}$}$ 2 M$_\odot$) to explain their properties \citep{howel06,hicken_07,yamanaka_09, Scalzo+2010, Tanaka+2010, Silverman+2011, Taubenberger+2011}. \cite{das_m12} explored the possibility of super-Chandrasekhar mass configurations arising from quantum mechanical modification of the degenerate electron equation of state in the presence of ultra strong internal magnetic fields.  The effect of this turns out to be sub-dominant to 
that of the Lorentz force; the latter by itself  can raise the maximum mass well above the Chandrasekhar limit \citep{Ostiker_Hartwick68, Bera+Bhattacharya2014}. The maximum mass of the magnetically supported configurations is dependent on the field geometry. Among  axisymmetric structures,  the maximum mass is about 1.9 M$_\odot$ for pure poloidal field \citep{Bera+Bhattacharya2014, das_m15, Franzon+Schramm2015} and more than 5 M$_\odot$ for pure toroidal field  \cite{Bera+Bhattacharya2016}, with intermediate values for mixed field configurations. These limits refer to equilibrium structures without consideration of stability.  In this paper, we study the stability of these equilibrium configurations.

Magnetic field is ubiquitous in the compact stars, be it neutron stars or white dwarfs. The highest magnetic field measured at the neutron star surface is $\sim 10^{15}$ G and that of a white dwarf is $\sim 10^9$ G \citep{Schmidt+2003}. While the field external to the star is primarily poloidal, \cite{Prendergast1956} suggests that in the interior both poloidal and toroidal fields must be present to ensure  long term stability. The consideration of the minimum energy principle \citep{Bernstein+1958} indicates that equilibrium configurations with pure poloidal \citep{marke73} or pure toroidal \citep{tayler1973} magnetic field are unstable. For such field geometries, perturbations in the matter and the magnetic field close to the neutral line (viz. the locus of vanishing magnetic field in the stellar interior, enclosed by field lines) can generate states of lower energy than the unperturbed configuration. Perturbation of the system would therefore drive it to a new configuration by rearranging
the magnetic field and matter. This magnetic instability is intrinsic to the field geometry, even if the magnetic energy is small compared to the thermal and the gravitational energy of the configuration (see e.g.  \cite{Flowers+Ruderman77}). \cite{tayler1973} and \cite{Acheson1978} show that for a pure toroidal configuration the non-axisymmetric azimuthal mode $m = 1$ is the dominant instability mode with very short instability time scale (Alfv\'en crossing time).

The non-linear evolution of the magnetic configurations with pure poloidal and pure toroidal field shows instability with growth time comparable to the Alfv\'en time of the configuration \citep{Braithwaite+Spruit2006, Braithwaite2006b, Bonanno+Uprin2013a, Bonanno+Uprin2013b, Bonanno+Uprin2013c, IbanezMejia+Braithwaite2015}. Configurations that show long term dynamical stability in numerical experiments with a stably stratified star contain comparable amounts of energy in poloidal and toroidal components \citep{Braithwaite+Nordlund2006, braithwaite09}. Numerically evolved axisymmetric or non-axisymmetric stable structures have been found for configurations with non-barotropic (stably stratified) equation of state \citep{mitchell+2015} and a helical initial field distribution with random or mixed poloidal-toroidal field \citep{Braithwaite2008}. 

Numerical studies of the evolution of neutron stars with pure poloidal fields in general relativistic formalism have been carried out by \cite{Lasky+2011, Ciolfi+2011, Ciolfi+Rezzolla2012}. These studies also indicate the presence of instability near neutral line.

Magnetars, a set of neutron stars with a strong surface magnetic field ($\sim 10^{15}$ G), show repeated gamma-ray flares. Quasi-periodic oscillations (QPOs) observed in the tail of giant flares provide evidence of neutron star oscillations \citep{Isreal+2005, Strohmayer+Watts2005}. Such oscillations of strongly magnetized neutron stars  have been investigated by various authors for both axisymmetric \citep{Glampedakis+2006, Lee2008, gabler+13, gabler+12} and non-axisymmetric \citep{Lander+2010, Lander+Jones2011a, Lander+Jones2011b, Asai+2015, Asai+2016} modes. 
 
In this paper, we study the linear and non-linear evolution of perturbed variables of a magnetized equilibrium structure. The perturbation equations and the methods of evolution are described in Section~\ref{eq+method}. In Section~\ref{results} we present  in brief the results obtained, which we discuss in Section~\ref{discussion}. Our conclusions are summarized in Section~\ref{conclusion}.

\section{Governing equations and Methods } \label{eq+method}

Here we model the self-gravitating, magnetized degenerate star with perfect conductivity in Newtonian gravity. The Newtonian description of gravity is adequate for this study as general relativity alters the white dwarf structure only by a few percent at maximum \citep{Bera+Bhattacharya2016}. We use a spherical polar coordinate system ($r$,~$\theta$,~$\phi$). The governing equations to describe the system are expressed as~:

\begin{align}\label{hydro_eqs}
\left(\frac{\partial\mathbf{v}}{\partial t}+(\mathbf{v}\cdot\nabla)\mathbf{v}\right) &= -\frac{1}{\rho}\mathbf{\nabla} P -\mathbf{\nabla} \Phi_g+\frac{1}{\rho}\left( \mathbf{J}\boldsymbol\times\mathbf{B}\right) \\
\frac{\partial\rho}{\partial t} &= -\nabla\cdot(\rho\mathbf{v})\\
\frac{\partial \mathbf B}{\partial t} &= \nabla\times(\mathbf{v}\times\mathbf{B})\\%-\textcolor{cyan}{\nabla\psi}\\
\mathbf{\nabla}^2\Phi_g &= 4\pi G \rho\\
P &= P(\rho)%k \rho^\gamma
\end{align}
Here, $\mathbf{v}, P, \rho, \Phi_g, \mathbf{J}, \mathbf{B}$ are the non-rotating part of the fluid velocity, pressure, matter density, gravitational potential, current and magnetic field respectively. The magnetic field components also satisfy divergence free condition $\mathbf{\nabla\cdot B} = 0$ and $\mathbf{\nabla\times B} = \mu_0\mathbf J$, $\mu_0$ being the free space permeability. The functional form of $P(\rho)$ describes the equation of state (EoS). Here we consider Fermi degenerate EoS to model the white dwarf and polytropic EoS ($P \propto \rho^{1+\frac{1}{n}}$, $n$ : polytropic index) in some cases to verify the results.

To study the evolution of the perturbed variables over the background equilibrium structure, initially the equilibrium configurations are constructed. These equilibrium  solutions without intrinsic fluid velocity are calculated from the time-independent part of the above equations.
\begin{align}\label{hydro_eui}
0 &= -\frac{1}{\rho_0}\mathbf{\nabla} P_0 -\mathbf{\nabla} {\Phi_g}_0+\frac{1}{\rho_0}\left( \mathbf{J_0}\boldsymbol\times\mathbf{B_0}\right) \\
\mathbf{\nabla}^2{\Phi_g}_0 &= 4\pi G \rho_0\\
P_0 &= P(\rho_0)%k \rho_0^\gamma
\end{align}
Here, the variables with the subscript zero represent the time independent or the equilibrium part of the variable. To solve for equilibrium axisymmetric magnetic configurations we follow the self-consistent field method \citep{hachi86, tomim05, lande09, Bera+Bhattacharya2014}. From the condition of axisymmetry we are restricted to only specific field geometries: pure poloidal, pure toroidal and mixed poloidal-toroidal to a certain extent.

After obtaining an equilibrium solution we add perturbation to it and aim to study the evolution of the variables. A radial perturbation generates stable oscillations when the effective polytropic index $n<3$ and this is generally satisfied in a white dwarf except for extreme relativistic configurations. Non-radial perturbations of the spherical star are classified as polar and axial depending on the parity. On the spherical surface polar perturbation can be decomposed in terms of $\hat rY_{lm}$ and $\nabla Y_{lm}$ and the axial perturbations into $\hat r\times\nabla Y_{lm}$. The polar perturbation generates $f,p$ mode oscillations with pressure as the restoring force and in a rotating star the axial perturbation generates $r$ modes from the Coriolis force \citep{Papaloizou+Pringle1978}. In a magnetic configuration axial perturbation excites the magnetic wave modes due to the presence of Lorentz force. Being interested in the evolution of the magnetic field we provide axial perturbations to the magnetic configurations. To keep the computation less expensive we assume that Cowling approximation is valid, i.e. the 
gravitational potential remains fixed to the initial condition. Cowling approximation is known to alter the frequencies of the acoustic waves (e.g., $f, g$), but the deviation is never more than 20\% in either Newtonian \citep{Cox1980} or relativistic \citep{Yoshida+Kojima1997} treatment.  Modes of higher order are affected less as the effect of oscillating components get averaged out. Although there is no direct study of the effect of the Cowling approximation on $r$ modes \citep{Kastaun2008} and magnetic modes, the approximation may be considered to be reliable as long as the perturbation does not significantly modify the equilibrium structure. In the following subsections, we describe the linear and non-linear methods followed to study the evolution of the perturbations to the equilibrium configurations.

\subsection{Linear perturbation }
To study the perturbations of the star in the linear regime we evolve the perturbed variables over the background equilibrium configuration. Such studies are common in the context of stellar pulsations of both non-magnetic \citep{Passamonti+2009, Jones+2002, Lockitch+Friedman1999, Papaloizou+Pringle1978} and magnetic \citep{Lander+2010, Lander+Jones2011a, Lander+Jones2011b, Asai+2015, Asai+2016} stars. The properties of the various modes of the oscillating stars are studied either using normal mode calculations or using MHD simulations. Here we are interested in investigating the axisymmetric strongly magnetized configurations which are generally deformed to either prolate or oblate shape depending on the field structure. Hence, we followed the $\phi$-decomposed MHD methods \citep{Lander+2010, Lander+Jones2011a, Lander+Jones2011b} to utilize the axisymmetry, and to deal with non-sphericity.

\subsubsection{Perturbed equations}
The general variables can be expressed as a sum of the equilibrium part and the perturbed part, i.e, $P = P_0+\delta P$, $\rho = \rho_0+\delta\rho$, $\mathbf{B} = \mathbf{B_0}+\delta\mathbf{B}$, $\mu_0\mathbf{J_0} = \nabla\times\mathbf{B_0}$, $\Phi_g={\Phi_g}_0$. The time dependent equations with terms of linear order in the perturbed variables can be expressed as, 

\begin{align}\label{hydro_pertb}
\frac{\partial\mathbf{f}}{\partial t} & = \left[-\mathbf{\nabla} \delta P + \frac{\mathbf{\nabla} P_0}{\rho_0}\delta\rho \right] \nonumber\\
& -\frac{1}{\rho_0}\left(\mathbf{J_0}\boldsymbol\times\mathbf{B_0}\right)\delta\rho +  \frac{1}{\rho_0}\left(\mathbf{J_0}\boldsymbol\times\boldsymbol{\beta}\right) \nonumber\\
& + \left[\frac{1}{\rho_0}\left(\nabla\times\boldsymbol{\beta}\right)\times\mathbf{B_0} - \frac{1}{\rho_0^2}\left(\nabla\rho_0\times\boldsymbol{\beta}\right)\times\mathbf{B_0}\right]\\
\frac{\partial\delta\rho}{\partial t} & = -\nabla\cdot\mathbf{f}\\
\frac{\partial \boldsymbol\beta}{\partial t} &= \nabla\times(\mathbf{f}\times\mathbf{B_0})-\frac{\nabla\rho_0}{\rho_0}\times(\mathbf{f}\times\mathbf{B_0})\\ %-\textcolor{cyan}{\rho_0\nabla\psi}\\
\delta P & = \left.\frac{\partial P}{\partial \rho}\right|_0\delta\rho \label{hydro_pertb_end}%\frac{\gamma P_0}{\rho_0} \delta\rho
\end{align} 
Here, $\mathbf{f}=\rho_0\mathbf{v}$ and $\boldsymbol\beta=\rho_0\delta\mathbf{B}$. Each of these perturbed quantities is further decomposed in azimuthal angle $\phi$ with index $m$, e.g. the perturbed pressure 
\begin{align}
 &\delta P (t, r, \theta, \phi) = \nonumber\\
 &\sum_{m=0}^{m=+\infty}[\delta P^+(t, r, \theta) \cos m\phi + \delta P^-(t, r, \theta) \sin m\phi].
\end{align}

This reduces the three-dimensional computation to a two-dimensional grid which helps limit the computation time.

\subsubsection{Initial values and Boundary conditions}
The magnetic equilibrium structure is calculated using the self-consistent field technique \citep{hachi86, tomim05, lande09, Bera+Bhattacharya2014}. Starting from an arbitrary initial configuration, a force balanced equilibrium configuration satisfying the virial condition is achieved iteratively. The solution is obtained for polytropic or Fermi degenerate equation of state. This force balanced solution is used as the background solution for perturbation study.

These magnetic equilibrium structures are non-spherical. In the linear perturbation study, to avoid difficulty in imposing boundary conditions on the non-spherical shape we use a modified radial coordinate $x=x(r, \theta)$, fitting the surfaces of constant pressure of the unperturbed star \citep{Jones+2002}. The partial differentials are expressed as
\begin{align}
 \left.\frac{\partial}{\partial r}\right|_\theta &= \left.\frac{\partial x}{\partial r}\right|_\theta \left.\frac{\partial}{\partial x}\right|_\theta,\\
 \left.\frac{\partial}{\partial \theta}\right|_r &= \left.\frac{\partial x}{\partial \theta}\right|_r \left.\frac{\partial}{\partial x}\right|_\theta + \left.\frac{\partial}{\partial \theta}\right|_x.
\end{align}
Here $x$ equals $r$ at a specific $\theta$ value $\theta = \theta_x$. We consider $\theta_x = \uppi/2 ~\text{and}~ 0$ for a prolate and oblate configuration respectively. The factors containing the partial derivative of $x$ with respect to $r$ and $\theta$ can be calculated at any point using the following relations,
\begin{align}
 \left.\frac{\partial x}{\partial r}\right|_\theta &= \left.\left.\frac{\partial \rho}{\partial r}\right|_\theta \right/\left.\frac{\partial \rho}{\partial x}\right|_{\theta_x},\\
 \left.\frac{\partial x}{\partial \theta}\right|_r &= \left.\left.\frac{\partial \rho}{\partial \theta}\right|_r \right/\left.\frac{\partial \rho}{\partial x}\right|_{\theta_x} .
\end{align}

At the stellar surface the Lagrangian pressure perturbation is zero i.e. for a displacement vector $\boldsymbol{\xi}$,
\begin{equation}
 \delta P + \boldsymbol{\xi}\cdot\nabla P_0 = 0.
\end{equation}
At the outer surface matter ($\rho_0$) and current ($\mathbf{J_0}$) densities vanish and Eq. $\ref{hydro_eui}$ provides us $\nabla P_0 = 0$. Therefore, at the outer surface $\delta P = 0$. Proper modeling of the atmosphere outside the stellar solid crust in order to study the numerical evolution of the field is difficult due to the very high value of the Alfv\'en velocity. As we are mainly interested in the evolution of the internal field, here we do not consider any evolution of the field or material outside the surface and we assume for simplicity that $\boldsymbol{\beta}=\mathbf{f}=0$ at the outer boundary. For the condition at the center we consider $\delta P(x=0) = 0$ $\boldsymbol{\beta}(x=0)=\mathbf{f}(x=0)=0$ as we plan to study the evolution of the non-axisymmetric modes (i.e. $m >0$). The perturbed variables also vanish at the pole for $m \neq 1$. For $m = 1$, $f_\theta, f_\phi,\beta_\theta ~\text{and}~ \beta_\phi$ are non-zero conserving their $\theta$-gradient.

Depending on the symmetry of the initial perturbation, perturbations of a specific class either axial or polar are excited. For the polar type of perturbation we use a spherical harmonic profile for initial density perturbations $\delta \rho \sim Y_{lm}(\theta, \phi)$. The axial type of perturbation is excited by introducing magnetic spherical harmonic perturbation $\mathbf f \sim \hat{r} \times \nabla Y_{lm}(\theta, \phi)$ in velocity.
\subsubsection{Numerical code} 
The set of $\phi$ decomposed linearly perturbed equations \ref{hydro_pertb}-\ref{hydro_pertb_end} are evolved forward in time from the initial condition mentioned above by using a MacCormack predictor-corrector algorithm \citep{MacCormack1969}. To dissipate spurious higher order oscillations, resulting from finite order grid differentiation, we impose additional fourth order Kreiss-Oliger dissipation. To ensure divergence free condition of the evolved perturbed field we use hyperbolic-parabolic divergence cleaning method \citep{Dedner+2002, Lander+2010}.

\begin{figure}
\centering
\includegraphics[width=0.47\textwidth]{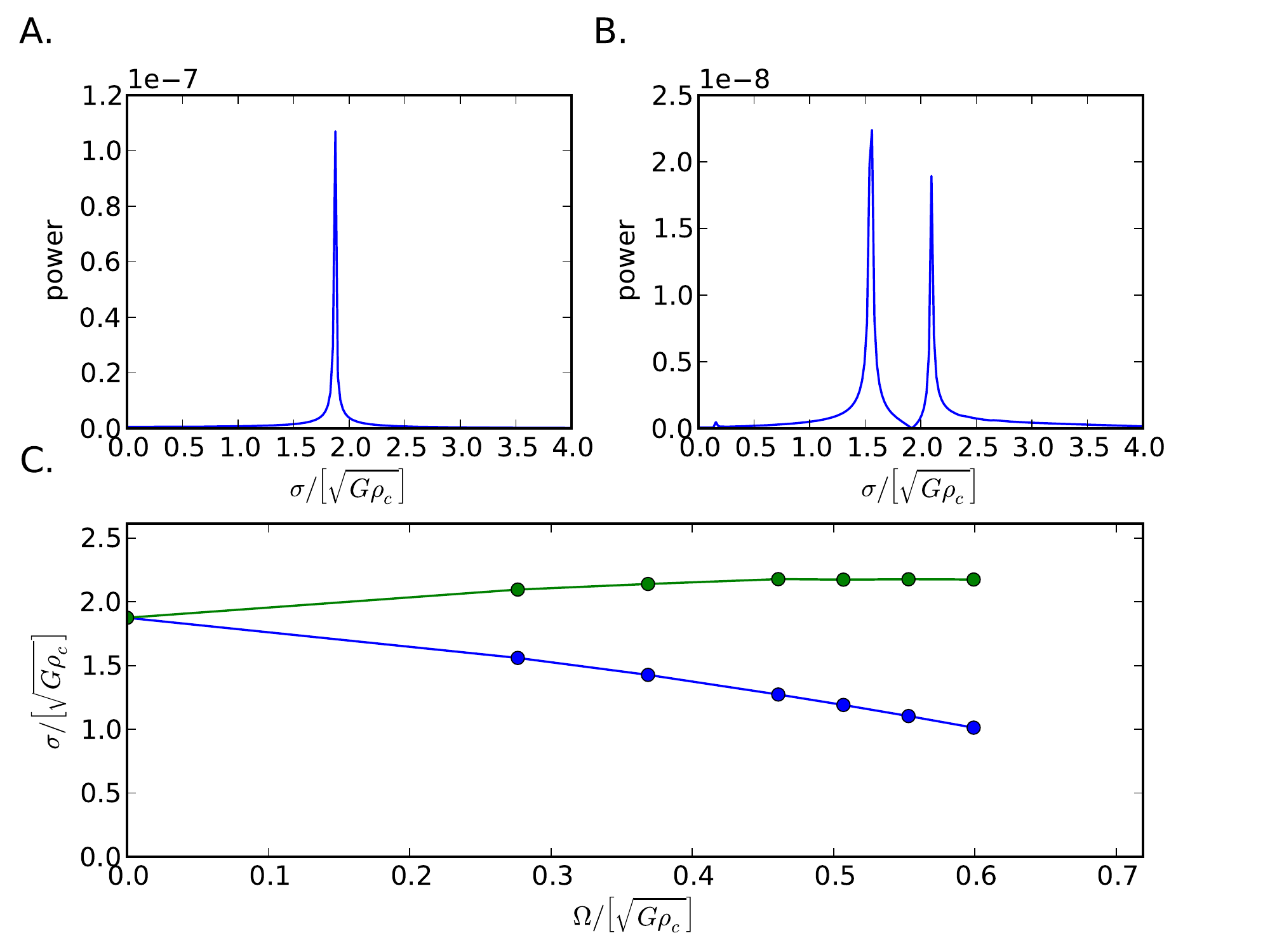}
\caption{A. Power spectra (in normalised unit) of $l=m=2$ $f$-mode oscillation from a polytropic star of polytropic index n = 1. B. Splitting of power spectra of a rotating star into co-rotating (lower frequency) and counter-rotating components. C. Dependence of mode frequencies on the rotation speed.}
\label{f_mode_power}
\end{figure}

To check the accuracy of the code we mainly rely on convergence tests. We provide the initial pressure perturbation as $\delta P \sim \rho \left(\frac{r}{R(\theta)}\right)^lY_{ll}(\theta, \phi)$ along with the equilibrium solution of a polytropic star of polytropic index n = 1 and study the $f-$mode oscillations for $l=m=2$. The perturbations evolve with a single peak frequency. But in the presence of rotation, it is split into two frequencies and the separation between them depends on the rotation speed. Fig.\ref{f_mode_power} show the $f-$mode oscillations of the rotating polytropic stars. These results are identical to the results reported in \citep{Jones+2002, Passamonti+2009}. Hence this verifies the performance of the non-magnetic parts of the code.

\begin{figure}
\centering
\includegraphics[width=0.47\textwidth]{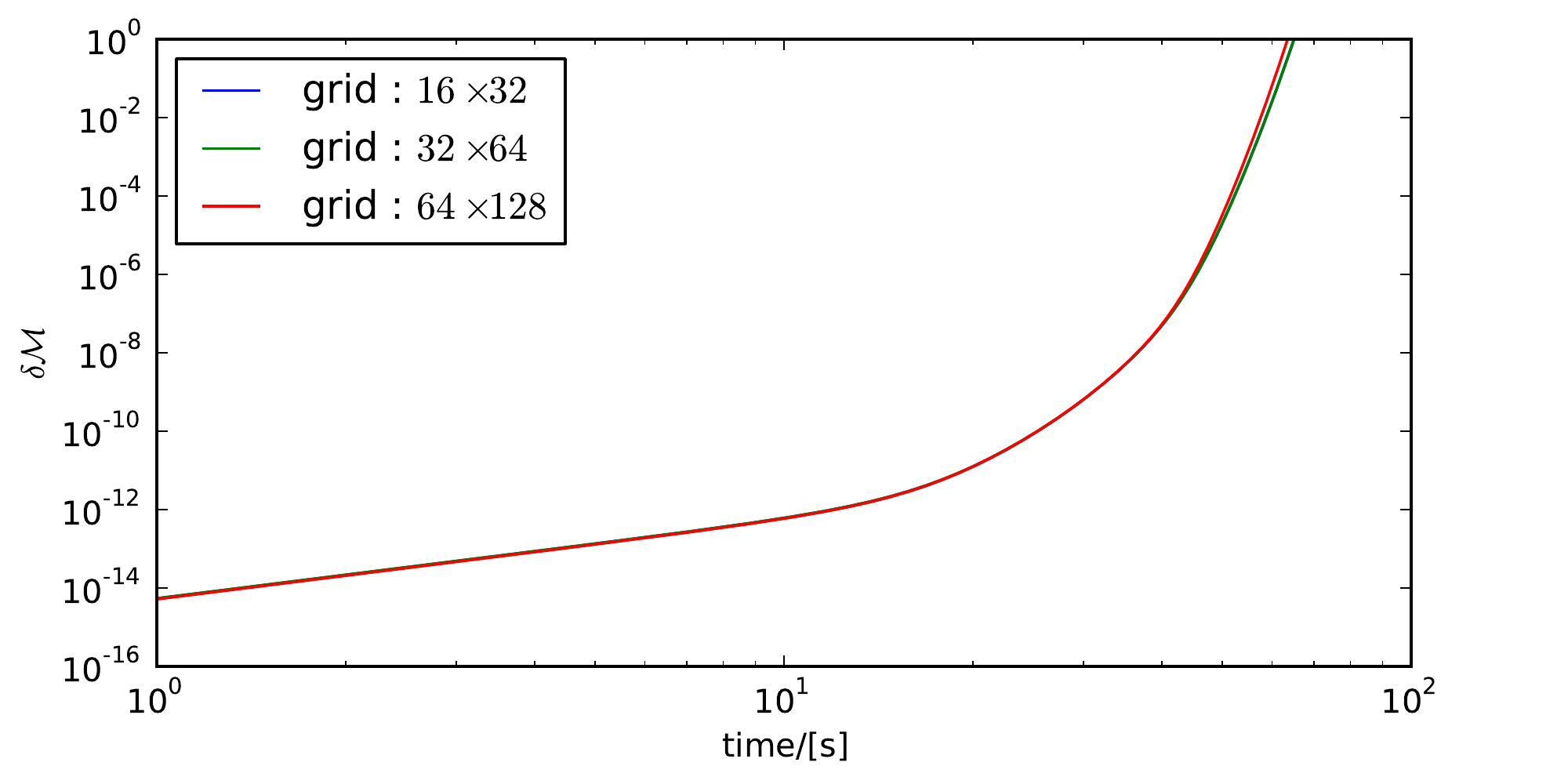}
\caption{ m = 1 instability of the magnetic degenerate star with a pure toroidal field. The perturbed magnetic energy (in normalised unit) increases with time. The behavior is independent of the grid resolution. 
}
\label{instability_accuracy}
\end{figure}

Magnetic configurations suffer from various kinds of instability depending on the field geometry. Here we proceed by choosing some of the known unstable modes and find the corresponding instability time scale. 
We study the evolution of the magnetic configuration for different grid resolutions and observe the appearance of instability. The initial perturbation is provided in the velocity terms $\mathbf f \sim \hat{r} \times \nabla Y_{lm}(\theta, \phi)$. Fig. \ref{instability_accuracy} suggests that the results are independent of the grid resolution and thus confirms the presence of instability.

\subsection{Nonlinear perturbation }
To study the evolution of the perturbation in the non-linear regime we use the MHD code $\textsc{pluto}$ which integrates a system of multidimensional conservation equations using Godunov-type shock-capturing schemes \citep{Mignone+2007}. This code is widely used to study various astrophysical problems related to accretion/protoplanetary disk, jet, outflow etc. We are interested in studying the evolution of the stellar internal field. Here also to avoid the modeling of the outer atmosphere we consider a spherical domain concentric to the stellar center and within the star. This spherical computation domain excludes the outer part near the surface with a very low density. We solve the MHD equations in $\textsc{pluto}$ using hll Riemann solver \citep{Harten+1983}. The divergence-free condition $\nabla\cdot \mathbf{B} = 0$ is enforced in the solution by coupling the induction equation to a generalized Lagrange multiplier \citep{Dedner+2002, Mignone+2010}. Here we consider the values of the parameters to be 
fixed 
to their initial 
values at the inner radial and impose the axisymmetric condition at the axial points. At the outer boundary, we maintain the fixed gradient of the equilibrium values such that the boundary allows the flow of matter \citep{Mukherjee+2013a}. This fixed gradient condition at the outer boundary may not suitable in a situation with strong inflow/outflow but provides an effective boundary for equilibrium with small perturbations. We evolve the perturbed polytropic star adiabatically with the adiabatic index $\gamma=\frac{5}{3}$. Here we assume the white dwarf as a polytropic star with polytropic index $n=1.5$ which is the non-relativistic approximation of the degenerate EoS. Hence for the non-linear evolution, we consider low mass ($<$M$_\odot$) magnetic white dwarfs where relativistic EoS does not modify the structure significantly.

In the non-linear evolution process, we compute perturbed quantities as the difference in the value of a variable at a given time from that in the equilibrium solution. These perturbed quantities are three dimensional variables. To identify non-radial modes we choose a $r$ value where the magnitude of the perturbation is close to maximum and  evaluate the $Y_{lm}$ coefficients of the $\theta-\phi$ plane data.

\section{Results } \label{results}
Here we present the results obtained from the linear and non-linear evolutions of the perturbation. We provide  specific perturbations to the equilibrium solution to study the instability. The non-linear stability is studied only for cases with pure poloidal and pure toroidal field assuming polytropic EoS whereas for all other cases we study the evolution using Fermi degenerate EoS with $\mu=2$ ($\mu$~: atomic mass per electron).

\subsection{Pure toroidal}

\subsubsection{linear perturbation}
\begin{figure}
\centering
\includegraphics[width=0.47\textwidth]{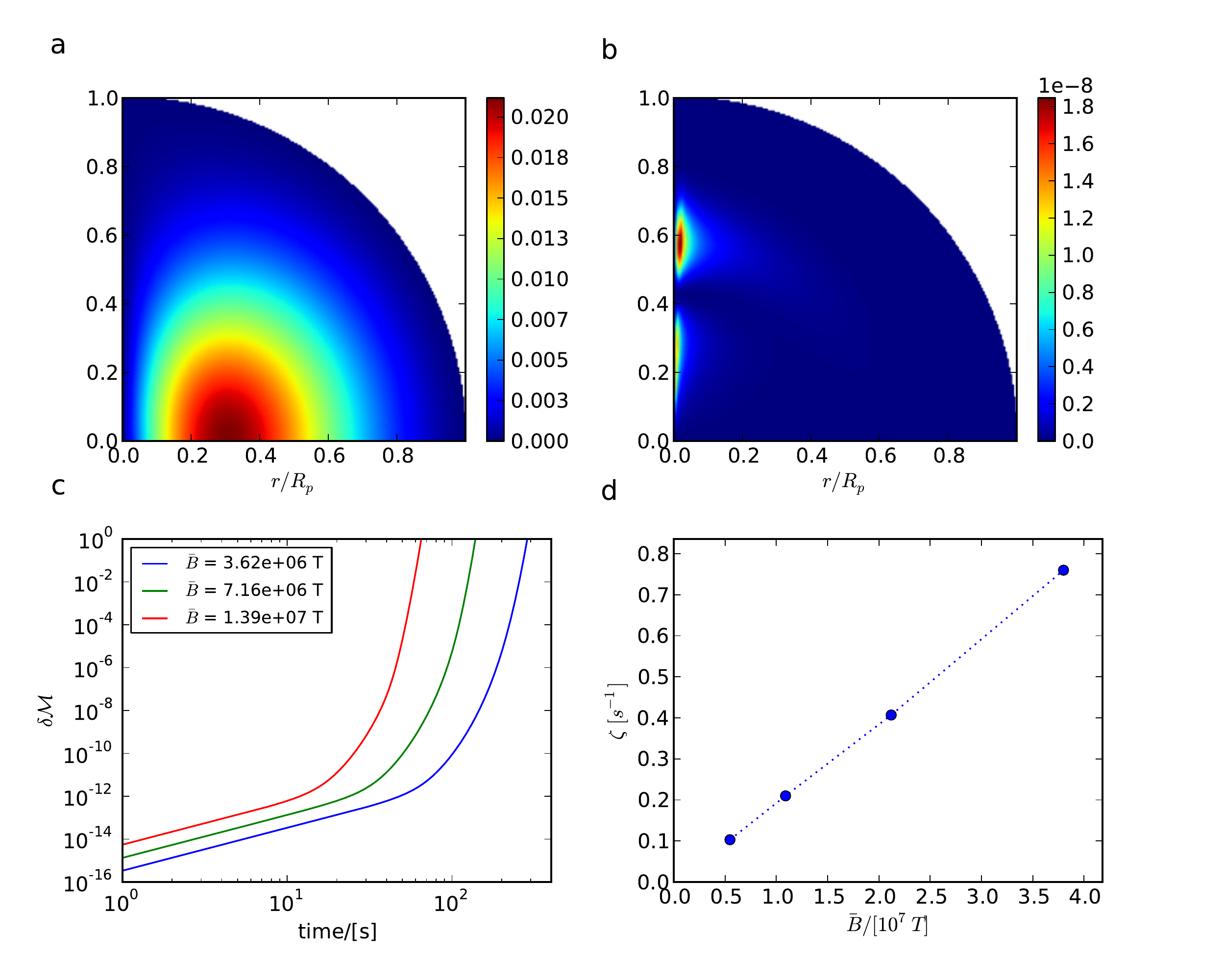}
\caption{Linear perturbation study of white dwarfs with pure toroidal field: a) Toroidal field distribution of the equilibrium configuration with mass 0.88 M$_\odot$ and $\mathcal{M}/W$=0.8\%. b) The ratio of perturbed field magnitude to the equilibrium field after the onset of the instability for the $m = 1$ evolution. The instability appears near the axial region of this configuration. c) The evolution of perturbed field shows instability which appears corresponding to their Alfv\'en crossing time $\tau_A\sim 78, 39 ~\rm{and}~ 20~ s$. d) The growth rate ($\zeta$), at the beginning of the instability from the time evolution of Fig.~\ref{lin_tor}c, is almost linearly proportional to the average magnetic field strength.}
\label{lin_tor}
\end{figure}

A configuration with pure toroidal field contains field only within the surface. The axial points have zero fields i.e. they form the magnetic neutral line (Fig.~\ref{lin_tor}a). Analytical perturbation study shows the presence of unavoidable non-axisymmetric (m=1) instability near the polar axis \citep{tayler1973}. To study this Tayler instability we consider the initial velocity perturbation as $\mathbf f \sim \hat{r} \times \nabla Y_{11}$. While the background field is only toroidal, the generated perturbed field contains all field components. The perturbed field energy increases with time linearly in the early phase but exponentially at later epochs (Fig.~\ref{lin_tor}c). The exponential increase in the perturbed field energy is the indication of instability. This being a linear calculation, the perturbation evolves over the background equilibrium solution and the total energy is not conserved in case of instability. Fig.~\ref{lin_tor}c shows that the appearance of the instability, i.e. the 
time when the magnetic energy starts to increase exponentially, is proportional to the Alfv\'en time ($\tau_A$). The Alfv\'en crossing time is the characteristic time of a magnetic configuration and is inversely proportional to the average field strength. This is estimated as $\tau_A\approx\frac{R}{\langle c_A\rangle} = R\sqrt{\frac{\mu_0\langle\rho\rangle}{\bar B^2}}$, where $R$ is the stellar radius, $\langle c_A\rangle$ is the volume averaged Alfv\'en speed. To quantify the growth rate of the instability we assume an exponential growth near the beginning of the instability and introduce a term $\zeta$, 
\begin{equation}
 \zeta = \frac{\Delta(\log\delta M)}{\Delta t}.
\end{equation}
Fig.~\ref{lin_tor}d shows that the instability growth rate is proportional to the average field strength. Therefore a strongly magnetized configuration with a pure toroidal field has a very short instability time scale. 

\subsubsection{non-linear perturbation}
\begin{figure}
\centering
\includegraphics[width=0.47\textwidth]{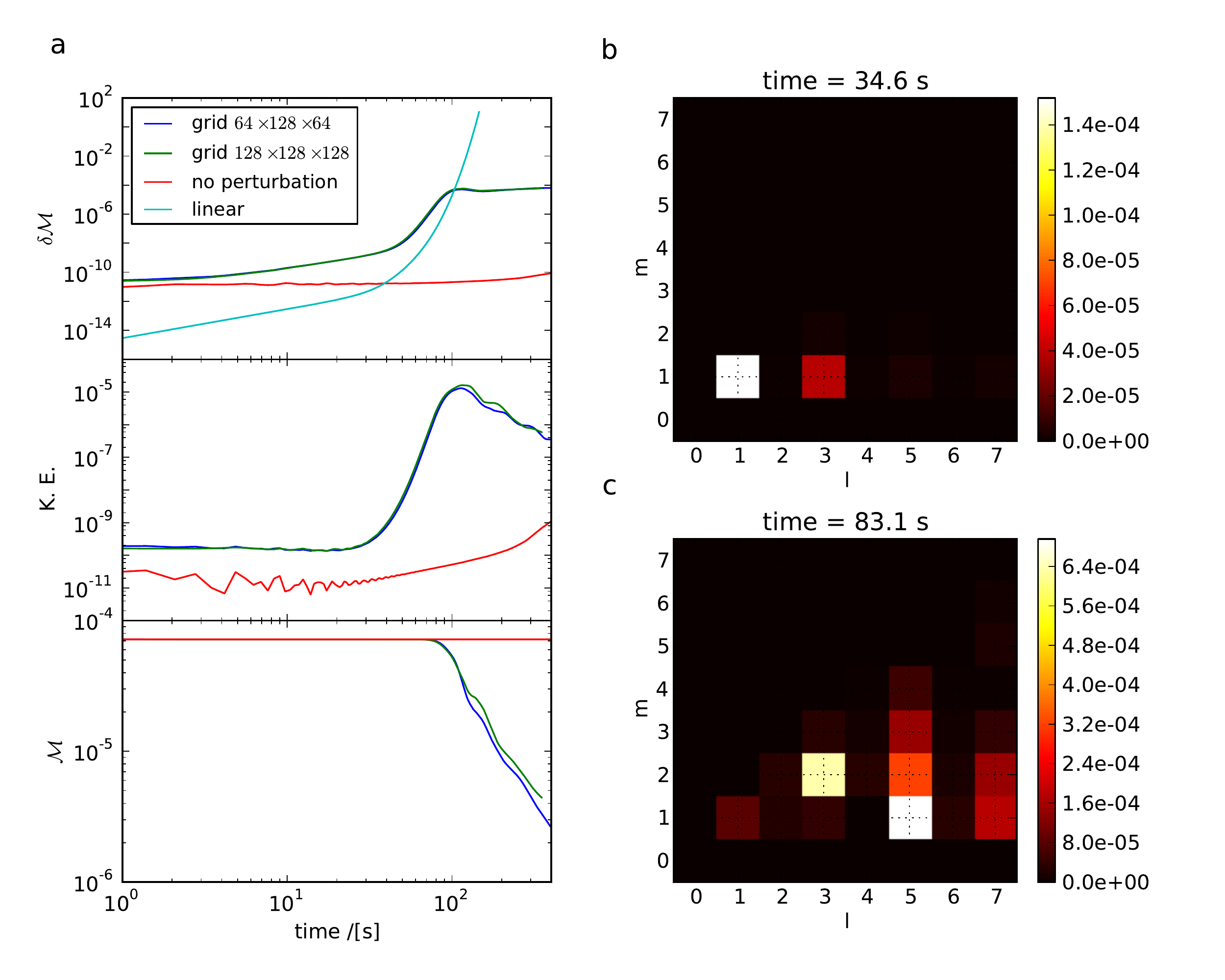}
\caption{Non-linear perturbation study of white dwarfs with a pure toroidal field: a) The evolution of i) perturbed magnetic energy ($\delta \mathcal M$), ii) kinetic energy (K.E.) and iii) total magnetic energy ($\mathcal M$). The results are almost independent of the grid resolution as the value of the parameters for $64^3$ and $128^3$ are identical. The linear evolution of the $m=1$ perturbation mode of this polytropic star is shown for comparison. The non-linear evolution of the equilibrium configuration without any perturbation shows the static characteristics. b) Early time (t~=~35~s) spherical harmonic mode components of the $\theta$-component of the perturbed field ($\delta B_\theta$) evaluated at $r/R=0.6$. c) Late time (t~=~83~s) mode components of $\delta B_\theta$ at the same $r$ value.}
\label{nonlin_tor}
\end{figure}

We study the non-linear evolution of the perturbed $\Gamma=\frac{5}{3}$ polytropic magnetic star with mass 0.78 M$_\odot$, radius 9859 km. The average field strength of this configuration is $6.1\times10^6$ T and the total magnetic energy is 0.8\% of the gravitational energy. Our computation region covers the range from 10\% to 90\% of the radius. The non-linear numerical evolution of this equilibrium solution (without any added perturbation) does not show any instability other than small fluctuations. For the perturbation study, axial perturbations as initial velocity $\mathbf v \sim \hat{r} \times \nabla Y_{22}(\theta, \phi)$ were added. The evolution of this perturbed configuration shows instability in the growth of the perturbed magnetic energy after about the Alfv\'en time $\tau_A \sim 35.7 ~s$ (Fig.~\ref{nonlin_tor}a). In the early phase, the perturbed field grows linearly before the instability. The time sequence plot of the total kinetic energy shows initial very minor change and 
then a sudden increase. This sudden increase in kinetic energy begins a little before the field growth which indicates that the enhanced velocity field induces the growth of the field. After the instability, the perturbed field magnitude shows saturation but the kinetic energy and the total magnetic energy decay significantly. This decay in magnetic energy may happen due to the assumed fixed gradient condition at the outer boundary which does not consider the whole configuration. Fig.~\ref{nonlin_tor}b shows that at the early time $l=m=1$ is the only dominating mode in the $\theta$-component of the perturbed field. At a later time, the amplitude of this specific mode increases, as well as other modes appear. To check the consistency of the results we study this non-linear evolution using $64^3$ and $128^3$ grid. The properties of the instability and other characteristics of the time evolution are nearly independent of the grid resolution (Fig.~\ref{nonlin_tor}a). We also study the linear 
evolution of $m=1$ mode perturbation for this polytropic configuration and observe the characteristics. Both the linear and non-linear study show consistent behavior in their time evolution. The main difference between the two is the non-appearance of the saturation in the linear case as there is no back reaction to modify the background solution.

\subsection{Pure poloidal}
\subsubsection{linear perturbation}
\begin{figure}
\centering
\includegraphics[width=0.47\textwidth]{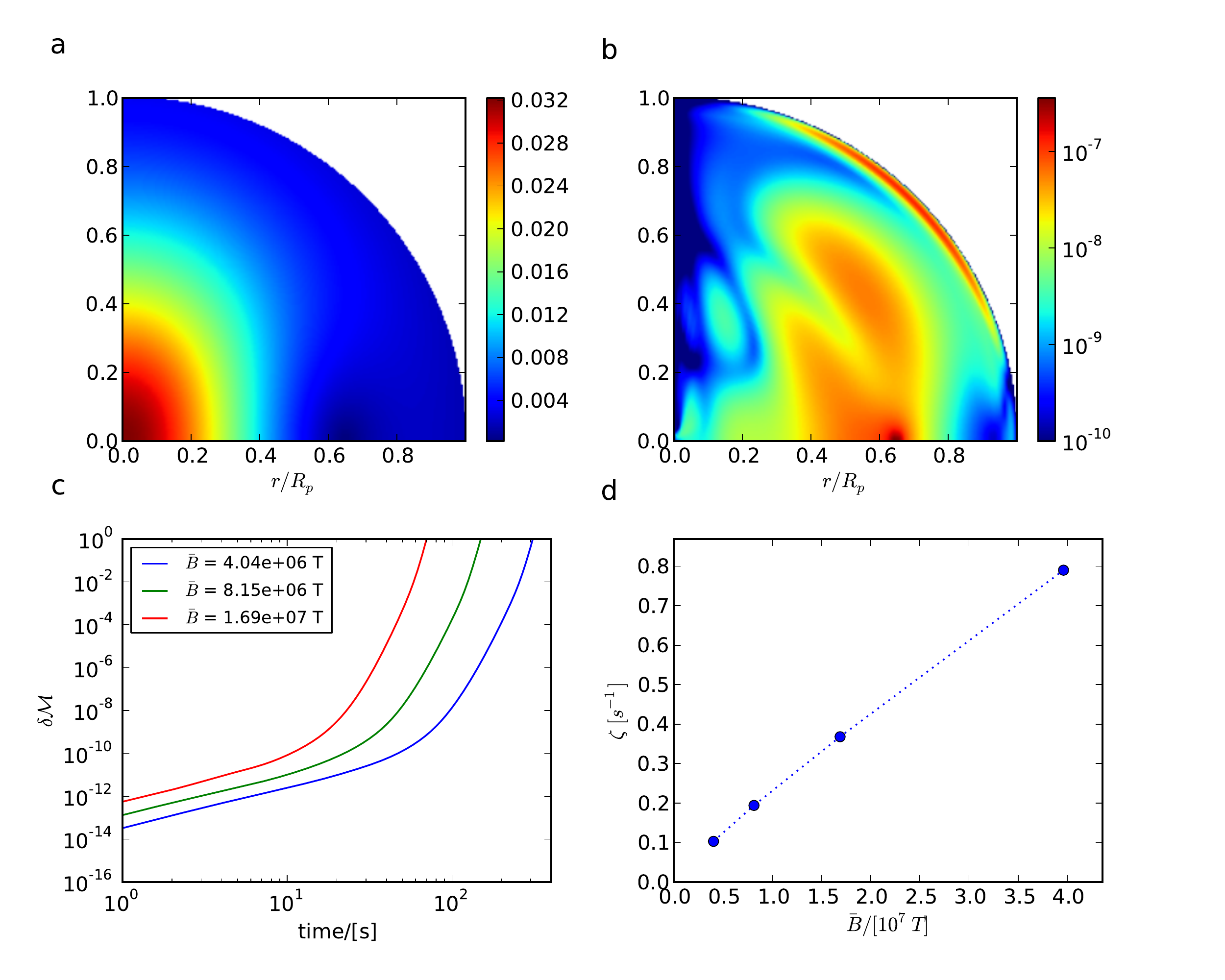}
\caption{ Linear perturbation study of white dwarfs with pure poloidal field (similar to Fig.~\ref{lin_tor}) : a) Poloidal field distribution of the equilibrium configuration with mass 0.88 M$_\odot$ and $\mathcal{M}/W$=0.8\%. b) The ratio of perturbed field magnitude to the equilibrium field after the onset of the instability for the $m = 2$ evolution. The instability appears in a region near the magnetic neutral line of the equilibrium configuration. c) The evolution of perturbed field shows instability which appears at the corresponding Alfv\'en crossing time $\tau_A\sim 69, 34 ~\rm{and}~ 17~ s$. d) The growth rate ($\zeta$), at the beginning of the instability from the time evolution of Fig.~\ref{lin_pol}c, is almost linearly proportional to the average magnetic field strength.}
\label{lin_pol}
\end{figure}

For a pure poloidal field configuration, the magnetic neutral line forms a circle on the equatorial plan. In the case of pure toroidal field structure with the magnetic neutral line along the axis, the perturbations evolve in the $\theta-\phi$ plane and we study $\phi$-decomposed mode. The perturbations generated from magnetic instability in a pure poloidal configuration will evolve in the $r-\theta$ plane. As we use our coordinate system with the origin at the stellar center, it is difficult to specify the mode of the perturbations from the symmetry. Here we study the evolutions of the $l=m=2$ mode of the perturbations providing the initial velocity perturbation as $\mathbf f \sim \hat{r} \times \nabla Y_{22}$. The time evolution of this perturbation on the equilibrium configuration (Fig.~\ref{lin_pol}A) with pure poloidal field exhibits a significant growth of the perturbed field near the neutral line (Fig.~\ref{lin_pol}B). Fig.~\ref{lin_pol}C indicates the appearance of the instability for the perturbed 
magnetic 
energy after the initial linear growth. Here also the exponential growth starts in corresponding Alfv\'en time. Fig.~\ref{lin_pol}D shows that the instability growth rate varies linearly with the average field strength of the configuration. 

\subsubsection{non-linear perturbation}
\begin{figure}
\centering
\includegraphics[width=0.47\textwidth]{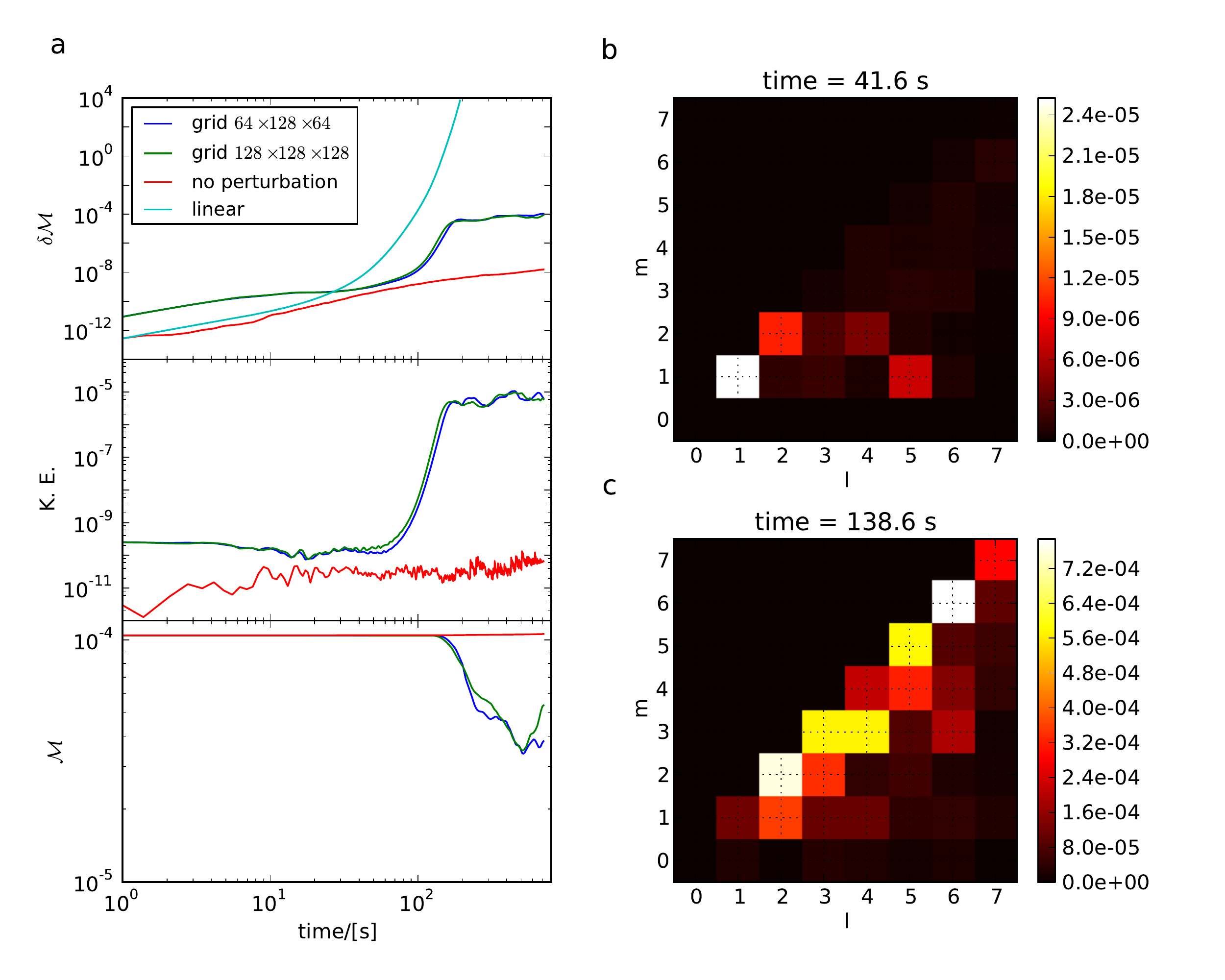}
\caption{Non-linear perturbation study of white dwarfs with pure poloidal field configuration (similar to Fig.~\ref{nonlin_tor}) : a) The evolution of i) perturbed magnetic energy ($\delta \mathcal M$), ii) kinetic energy (K.E.) and iii) total magnetic energy ($\mathcal M$). The linear evolution is done for $m=2$ mode of the perturbation. b) Early time (t~=~42~s) spherical harmonic mode components of the $\phi$-component of the perturbed field ($\delta B_\phi$) evaluated at $r/R=0.6$. c) Late time (t~=~139~s) mode components of $\delta B_\phi$ at the same $r$ value.}
\label{nonlin_pol}
\end{figure}

Similar to the pure toroidal case here too we consider a $\Gamma=\frac{5}{3}$ polytropic star which has a mass 0.9 M$_\odot$, radius 9716 km and average magnetic field of strength $7\times 10^6$~T. For this configuration, the magnetic field energy is 0.9\% of the gravitational energy and the Alfv\'en crossing time $\tau_A \sim 32 ~s$. The non-linear evolution of the equilibrium structure almost remains unchanged when the variables on the spherical shell ($0.1\leq r/R\leq0.9$) are evolved using $\textsc{pluto}$. For the perturbation evolution study we add the axial velocity perturbation $\mathbf v \sim \hat{r} \times \nabla Y_{22}(\theta, \phi)$ to the equilibrium solution and let it evolve. Now, the perturbed magnetic field energy and the kinetic energy exhibit instability after about a few Alfv\'en crossing time (Fig.~\ref{nonlin_pol}a). It is also observed that after these values attain some kind of saturation the total magnetic energy decays, indicating
extreme non-linear interactions and deformation of matter flow and field geometry. The Newtonian evolution properties ( e.g. appearance of instability, significant magnetic field energy decay within a few Alfv\'en times after the beginning of instability, generation of toroidal field energy and its rise to a level comparable that of the poloidal field) match with the general relativistic evolution of the poloidal neutron star explored by \cite{Ciolfi+Rezzolla2012}. The linear study of the $m=2$ mode also shows similar instability characteristics but in this case the instability happens a little earlier in comparison to the non-linear case. This may indicate that the particular mode (i.e. $m=2$) used to study linear evolution may not be the dominant one. The mode decomposition of the $\phi$-component of the perturbed field shows initially $m=1$ or $m=2$ modes (Fig.~\ref{nonlin_pol}b) but later many other modes arise (Fig.~\ref{nonlin_pol}c).

\subsection{Effects of Mixed field, Rotation} \label{mixed+rotation}
 In the last two subsections, we observed that the pure toroidal and pure poloidal field configurations are unstable as they show instability when we perturb the system. Both of these field configurations, assumed here for the perturbation study, are idealisations. We may expect more realistic magnetic configurations to contain both poloidal and toroidal components without any specific symmetry. On the other hand, the intrinsic spin of the star may play a role in aligning the magnetic field axis. Although many such complex possibilities exist, our investigation in this paper remains restricted only to axisymmetric configurations.
 
\begin{figure}
\centering
\includegraphics[width=0.47\textwidth]{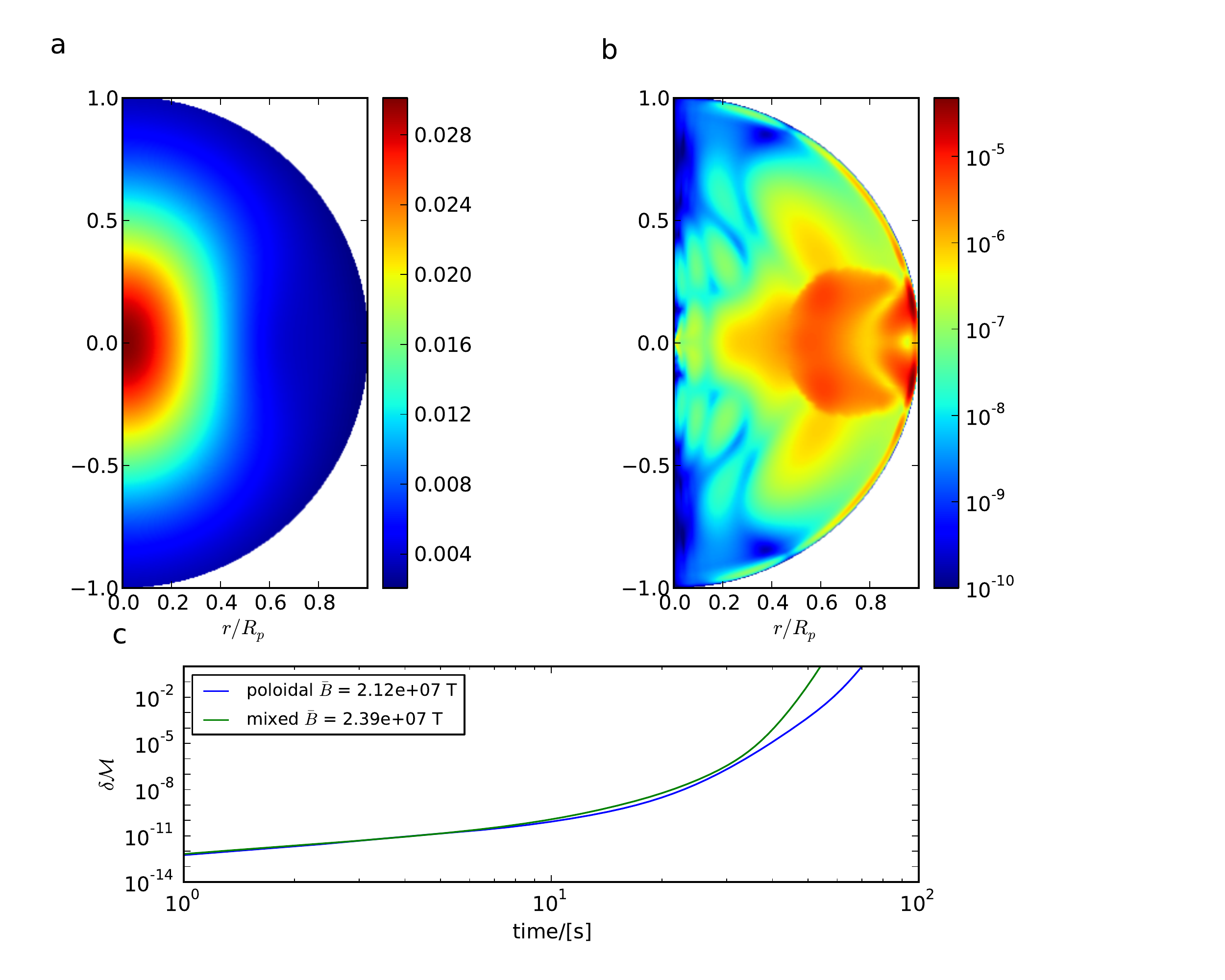}
\caption{Linear instability of a mixed field configuration. a) Equilibrium field distribution is mainly dominated by the poloidal field. The toroidal field component is there within the closed field line bounded by the stellar surface. b) The distribution of the perturbed field compared to the equilibrium field strength. c) The time variation of the perturbed field magnitude shows instability similar to the pure poloidal field configuration.}
\label{lin_mix}
\end{figure}

As mentioned above, the axisymmetric equilibrium configurations we deal with are computed using the self-consistent field method. One of the limitations of this method is that one can not obtain a configuration with comparable poloidal and toroidal field. The maximum energy of the toroidal component in the mixed field configuration is limited to less than 10\% \citep{armaza+15} of the total. Although \cite{Ciolfi+Rezzolla2013} found mixed field configurations with a significant toroidal component using a perturbative method, we do not find such solutions to represent equilibrium structures. Here we study the linear evolution of the perturbation of a mixed field configuration containing about 2\% of the magnetic energy in the toroidal form (Fig.~\ref{lin_mix}A). As the total energy in the toroidal component is not significant compared to that in the poloidal component we find the magnetic instability characteristics to be similar to the pure poloidal case (Fig.~\ref{lin_mix}C). However, in an equatorial region near the stellar surface, where the local toroidal field significantly exceeds the local poloidal component, a change in the nature of the instability is observed. A comparison of Fig.~\ref{lin_pol}B and Fig.~\ref{lin_mix}B illustrates this difference. At the boundary of the toroidal field region the poloidal current distribution displays a discontinuity. Potential numerical errors associated with this restricts our present study to configurations with relatively low toroidal field. The location of the region being very close to the surface of the star also makes it difficult to define a \textsc{pluto} computation box sufficiently in the interior. Hence for this case we restrict ourselves to linear perturbation study.
 
\begin{figure}
\centering
\includegraphics[width=0.47\textwidth]{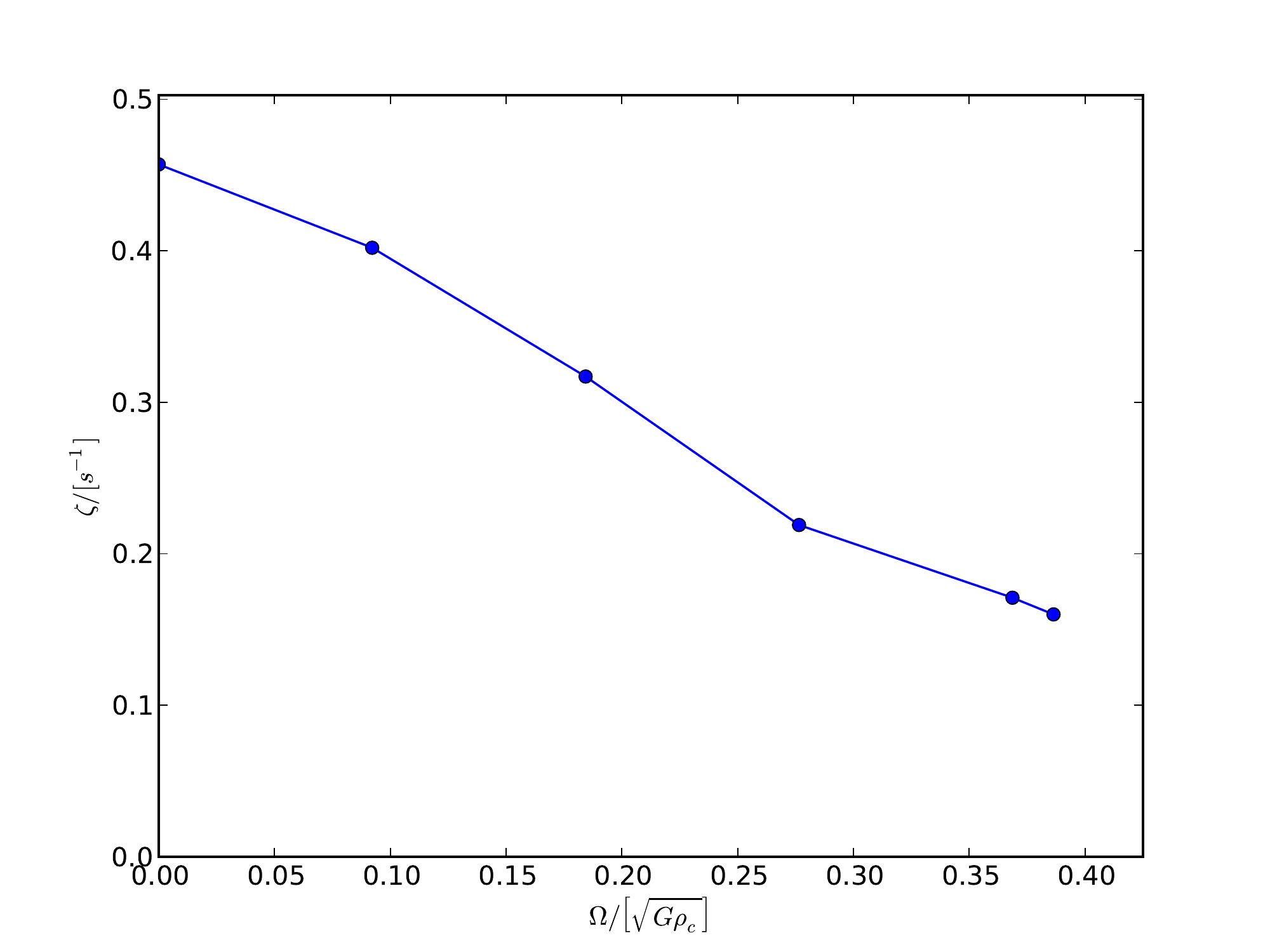}
\caption{The growth rates ($\zeta$) of the rotating 0.88 M$_\odot$ white dwarfs with pure toroidal field ($|\mathcal M/W| = 0.8$\%). The maximum rotation frequency presented here is close to its Keplerian frequency. The growth rate reduces by more than 50\% as the rotation frequency increases. }
\label{tor_rot_growth_rate}
\end{figure}

Axisymmetric configurations with uniform rotation along the symmetry axis can be obtained by introducing the centrifugal force in the equilibrium force balance equation.The centrifugal force acts outward from the rotation axis and is proportional to the square of the angular frequency. In the presence of this force, one can get an equilibrium solution upto a frequency, known as Keplerian frequency, beyond which a bound object cannot be formed. To study the perturbation of these configurations one must consider the Coriolis force term in the perturbation equations. The presence of Coriolis force provides $r$-mode oscillations for a non-magnetic star and reduces the instability growth rates of the magnetic configurations. Fig.~\ref{tor_rot_growth_rate} shows the variation of the growth rate depending on the rotation frequency of a magnetic white dwarf of mass 0.88 M$_\odot$ and $|\frac{\mathcal M}{W}| = 0.8$\%. Here from the 
linear study we observe that the instability growth rate decreases by more than 50\% as the rotation speed approaches Keplerian frequency.

\subsection{Super-Chandrasekhar mass white dwarfs}
\begin{figure}
\centering
\includegraphics[width=0.47\textwidth]{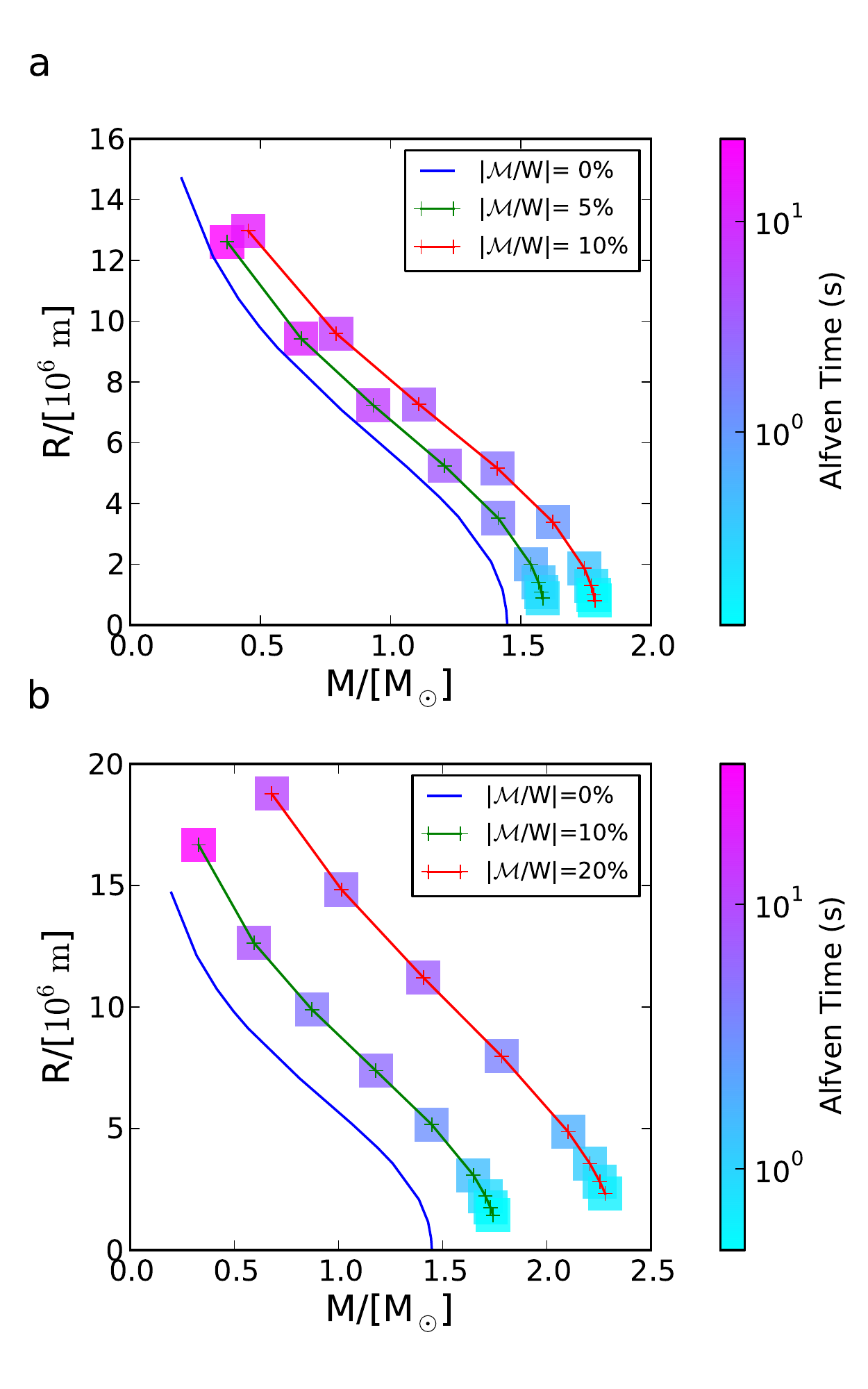}
\caption{The mass-radius relation of white dwarfs with pure poloidal (upper panel) and pure toroidal field. The maximum mass for a pure poloidal field is about 1.9 M$_\odot$ and for pure toroidal field it is more than 5 M$_\odot$. The coloured squares exhibit the Alfv\'en crossing time ($\tau_A$) of these configurations.}
\label{MR_alfven_time}
\end{figure}

 The presence of strong ordered magnetic field can deform the equilibrium structure. Depending on the field geometry and hence the direction of Lorentz force the equilibrium structure can be prolate or oblate. As the Lorentz force becomes significant it can support more mass relative to the non-magnetic configuration. Fig.~\ref{MR_alfven_time} shows that the mass-radius relations of the magnetic white dwarfs with equal $|\mathcal M/W|$ ratio are shifted to the higher mass for both the pure poloidal and pure toroidal field geometry. The magnetically supported super-Chandrasekhar mass white dwarfs have very short Alfv\'en crossing time scales, typically less than a second.

\begin{figure}
\centering
\includegraphics[width=0.47\textwidth]{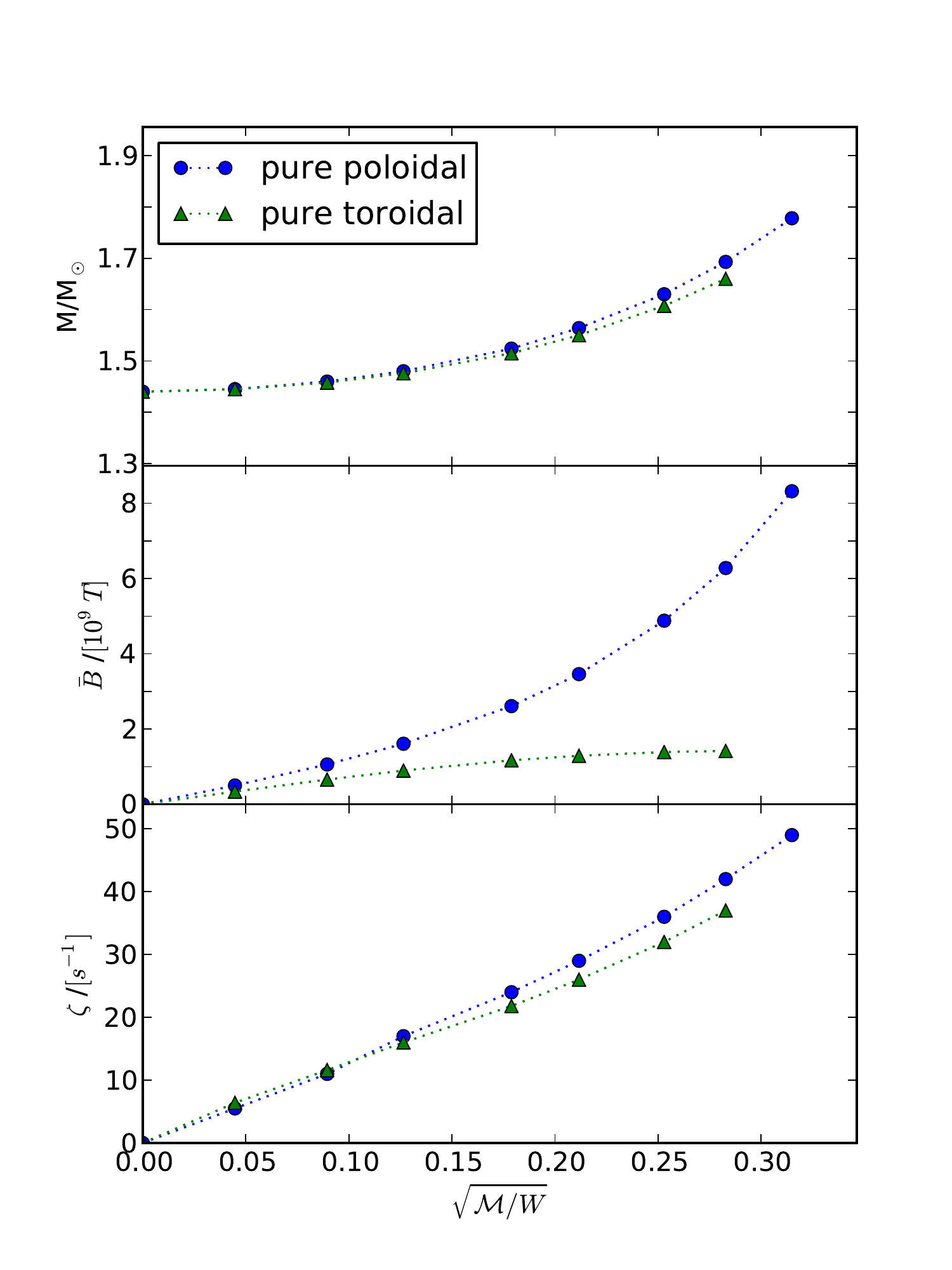}
\caption{The effective magnetic field ($\sqrt{|\mathcal M/W|}$) dependency of the configuration mass, average magnetic field and the instability time scale of supar-Chandrasekhar mass white dwarfs with central density $2\times 10^{13}~ \rm{kg/m}^3$.}
\label{lin_SuCh}
\end{figure}

We study the linear evolution of perturbation in magnetized super-Chandrasekhar mass white dwarfs with central density $2\times 10^{13}~ \rm{kg/m}^3$. Fig.~\ref{lin_SuCh} shows that the equilibrium mass increases as the ratio of magnetic to the gravitational energy of the configuration increases, irrespective of the field geometry -- whether pure poloidal or pure toroidal. 

The instability growth rates for these configurations vary almost linearly with the effective magnetic field strength, defined as the square-root of the ratio between magnetic to gravitational energy ($\sqrt{|\mathcal{M}/W|}$). This behaviour is akin that obtained using average field in configurations with field strengths too low to affect the stellar structure (Fig.~\ref{lin_tor}d \& Fig.~\ref{lin_pol}d) . The non-linear evolution of these perturbed highly massive magnetic configurations is expected to show instability behaviour similar to those presented above for low-mass structures. We do not attempt to study the adiabatic non-linear evolution of the perturbation as the full Fermi degenerate equation of state of super-Chandrasekhar mass white dwarfs has a varying polytropic index. The linear evolution growth rates are presented for magnetic configurations with $|\mathcal{M}/W|$ value upto to $\sim$10\%. Beyond this the value is highly dependent on the grid resolution. The obtained growth rates of these cases 
from the linear study are very high ($>10~s^{-1}$), indicating the highly unstable nature of these configurations.

\section{Discussion } \label{discussion}
In this perturbation study, we consider ideal MHD with no viscosity effects. Fields with long range order as considered here cause white dwarfs to show instability with sub-second time scales. For a typical white dwarf parameter, the viscous dissipation time is about $10^{10}$~yr and so is the resistive time scale \citep{Bera+Bhattacharya2016}. Hence, the characteristics of the instability are not significantly affected by these dissipative effects in the early stage. The presence of high-speed rotation, in general, reduces the instability growth rate of magnetic structures \citep{Acheson1978, Pitts+Tayler1985, Braithwaite2006a, 
Lander+Jones2011a, Lander+Jones2011b}, as shown in Section~\ref{mixed+rotation} for the specific case of  toroidal fields. White dwarfs with higher central density and higher field strength also have higher Keplerian frequency \citep{Franzon+Schramm2015}, suggesting that rotation may provide a stabilizing influence. However it can not stabilize the system completely unless the toroidal field increases very slowly with the distance from the axis \citep{Bonanno+Uprin2013c}. The toroidal field increases faster near the axis for more massive magnetic white dwarfs and this ensures that even the rotating configurations remain unstable.

Another idealization considered in this work is the zero-temperature degenerate equation of state. At the very outer layers near the surface of the star, matter may however be non-degenerate and thus have a different equation of state. This is not likely to impact on the nature of the instability shown here since the driving mechanism operates in deeper layers. In fact results for a degenerate gas differ little from those for a polytropic one, suggesting near independence on the equation of state. Long-term evolution of the instability and observables such as the luminosity may, however, be affected more significantly by the state of the outermost layers. The computation of these effects is beyond the scope of this paper.

The state of the configuration after the instability is not studied here but we observe that during non-linear evolution many large-amplitude fluctuating components with significant matter flow are created as the instability grows. The ultimate destiny of the evolved state will be decided on whether it is able to settle into a configuration in dynamic equilibrium. So far such states of dynamic equilibrium have been found only in cases with magnetic energy too low to influence the stellar structure \citep{Braithwaite+Nordlund2006, braithwaite09, Ciolfi+2011}. This indicates that while low field configurations may achieve a dynamical equilibrium state, the possibility is remote for this occurring in strongly magnetized configurations where the instability is mainly driven by the electrical currents. Our numerical study suggests that the strongly magnetized configurations have an extremely complex evolution which may lead to an eventual collapse of the system as these configurations are already very close to 
the maximum mass limit. 

However, the formation scenarios of white dwarfs also leave little room for generating configurations supported by ultra strong magnetic fields. Strong magnetic fields in white dwarfs are thought to be generated from either i) inheritance from the progenitor star via flux-freezing \citep{Ruderman1972}, or, ii) possible dynamo action during the common envelope phase of a binary system \citep{Potter+Tout2010, Briggs+2015}. White dwarfs are formed by gravitational contraction of stars less massive than $\sim$ 8 M$_\odot$ in a process lasting several days \citep{Woosley+Weaver1986}. If an ultra strong field is inherited from flux-freezing, then the instability discussed above would prevent the ordered field structure from being retained until the formation of the white dwarf is complete. On the other hand, the dynamo mechanism during the common envelope phase is a steady process which takes about a Myr \citep{Potter+Tout2010} to produce the field at the white dwarf surface. Instabilities with short time scales 
will then cause the field, if it is ultra strong,  to evolve to a more stable configuration which may not possess a significant long range order.  In absence of such long range order the effective Lorentz force would be insignificant and will be incapable of supporting super-Chandrasekhar mass configurations.

\section{Conclusions} \label{conclusion}
In this paper, we have studied the linear and non-linear evolution of perturbations to an axisymmetric, strongly magnetized object.  Our main results are:

\begin{enumerate}
\item Axisymmetric magnetic configurations with pure poloidal or pure toroidal field suffer from magnetic instabilities with time scale comparable to the Alfv\'en crossing time ($\tau_A$) of the configuration. As the Alfv\'en crossing time is inversely proportional to the average  magnetic field of the configuration, structures with ultra strong magnetic fields, with very short Alfv\'en crossing times, are strongly unstable.
\item In the case of rotating magnetic white dwarfs, the instability growth rate reduces as the rotation speed increases. However this is insufficient to fully stabilize magnetically supported configurations near their mass limits.
\item Magnetically supported super-Chandrasekhar mass white dwarfs require extremely strong magnetic fields in the interior and are hence susceptible to instabilities with a very short growth time scale (typically less than a second). Instabilities of this nature may in fact prevent the formation of such objects.
\end{enumerate}

\section{Acknowledgement}
PB thanks CSIR, India for Research Fellowship grant SPM-09/545(0221)/2015-EMR-I. We thank E. Truhlik, H. Spruit and J. P. Ostriker for the useful comments. We also thank the referee for the valuable comments that helped us to improve the paper. The most of the numerical computations were carried out using IUCAA HPC.
%==========================================
%----------------- Bibliography and bibfile

\def\aj{AJ}%
\def\actaa{Acta Astron.}%
\def\araa{ARA\&A}%
\def\apj{ApJ}%
\def\apjl{ApJ}%
\def\apjs{ApJS}%
\def\ao{Appl.~Opt.}%
\def\apss{Ap\&SS}%
\def\aap{A\&A}% 
\def\aapr{A\&A~Rev.}%
\def\aaps{A\&AS}%
\def\azh{AZh}%
\def\baas{BAAS}%
\def\bac{Bull. astr. Inst. Czechosl.}%
\def\caa{Chinese Astron. Astrophys.}%
\def\cjaa{Chinese J. Astron. Astrophys.}%
\def\icarus{Icarus}%
\def\jcap{J. Cosmology Astropart. Phys.}%
\def\jrasc{JRASC}%
\def\mnras{MNRAS}%
\def\memras{MmRAS}%
\def\na{New A}%
\def\nar{New A Rev.}%
\def\pasa{PASA}%
\def\pra{Phys.~Rev.~A}%
\def\prb{Phys.~Rev.~B}%
\def\prc{Phys.~Rev.~C}%
\def\prd{Phys.~Rev.~D}%
\def\pre{Phys.~Rev.~E}%
\def\prl{Phys.~Rev.~Lett.}%
\def\pasp{PASP}%
\def\pasj{PASJ}%
\def\qjras{QJRAS}%2215.bib
\def\rmxaa{Rev. Mexicana Astron. Astrofis.}%
\def\skytel{S\&T}%
\def\solphys{Sol.~Phys.}%
\def\sovast{Soviet~Ast.}%
\def\siamr{SIAMR}%
\def\ssr{Space~Sci.~Rev.}%
\def\zap{ZAp}%
\def\nat{Nature}%
\def\iaucirc{IAU~Circ.}%
\def\aplett{Astrophys.~Lett.}%
\def\apspr{Astrophys.~Space~Phys.~Res.}%
\def\bain{Bull.~Astron.~Inst.~Netherlands}%
\def\fcp{Fund.~Cosmic~Phys.}%
\def\gca{Geochim.~Cosmochim.~Acta}%
\def\grl{Geophys.~Res.~Lett.}%
\def\jcp{J.~Chem.~Phys.}%
\def\jgr{J.~Geophys.~Res.}%
\def\jqsrt{J.~Quant.~Spec.~Radiat.~Transf.}%
\def\memsai{Mem.~Soc.~Astron.~Italiana}%
\def\nphysa{Nucl.~Phys.~A}%
\def\physrep{Phys.~Rep.}%
\def\physscr{Phys.~Scr}%
\def\planss{Planet.~Space~Sci.}%
\def\procspie{Proc.~SPIED}%
\let\astap=\aap
\let\apjlett=\apjl
\let\apjsupp=\apjs
\let\applopt=\ao

\bibliographystyle{mnras}	% (uses file "plain.bst")

\bibliography{ref.bib}
\label{lastpage}
\end{document}